\documentclass[preprint]{revtex4-2}
\usepackage{amsmath,amssymb,amsfonts}
\usepackage{amsmath,amssymb,amsfonts}
\usepackage{graphicx}
\usepackage{geometry}
\usepackage{fancyhdr}
\usepackage{citesort}
\usepackage{xcolor}
\usepackage{soul}
\usepackage{physics}
\usepackage{hyperref}
\usepackage{xr-hyper}
\usepackage{lineno}
\AtBeginDocument{\thispagestyle{fancy}}  
\begin{document}
\title{
Shear induced topological changes of local structure in dense colloidal suspensions\\
}
\author{Ratimanasee Sahu$^{1}$, Abhishek Kumar Gupta$^{1,2}$, Peter Schall$^{2}$, Sarika Maitra Bhattacharyya$^{3,4}$,\textsuperscript{*}, Vijayakumar Chikkadi$^{1}$\textsuperscript{$\dagger$}}

\affiliation{
$^1$ Physics Division, Indian Institute of Science Education and Research Pune, Pune-411008, India.\\
$^2$ Institute of Physics, University of Amsterdam, 1098 XH Amsterdam, The Netherlands.\\
$^3$ Polymer Science and Engineering Division, CSIR-National Chemical Laboratory, Pune-411008, India.\\
$^4$ Academy of Scientific and Innovative Research (AcSIR), Ghaziabad 201002, India\\
}

\begin{abstract}
Understanding the structural origins of glass formation and mechanical response remains a central challenge in condensed matter physics. Recent studies have identified the local caging potential experienced by a particle due to its nearest neighbors as a robust structural metric that links microscopic structure to dynamics under thermal fluctuations and applied shear. However, its connection to locally favored structural motifs has remained unclear. Here, we analyze structural motifs in colloidal crystals and glasses and correlate them with the local caging potential. We find that icosahedral motifs in glasses are associated with deeper caging potentials than crystalline motifs such as face-centered cubic (FCC) and hexagonal close-packed (HCP) structures. Both crystalline and amorphous systems also contain large number of particles belonging to stable defective motifs, which are distortions of the regular motifs. Under shear, large clusters of defective motifs fragment into smaller ones, driving plastic deformation and the transition from a solid-like to a liquid-like state in amorphous suspensions. Particles that leave clusters of stable motifs are associated with shallower caging potentials and are more prone to plastic rearrangements, ultimately leading to motif disintegration during shear. Our results thus reveal that the loss of mechanical stability in amorphous suspensions is governed by the topological evolution of polytetrahedral motifs, uncovering a structural mechanism underlying plastic deformation and fluidization.
\end{abstract}

\maketitle

\section*{Introduction}

Amorphous solids such as glasses exhibit short-range structural order, in contrast to crystals, which possess long-range periodic order. The structural organization in disordered solids has been studied through a variety of approaches, including two-point density correlations \cite{Dullens14, Torquato07, Bhattacharyya17, Maia17}, free volume analysis \cite{Cohen61, Grest79, Spaepen77}, bond-orientational order (BOO) parameters \cite{Ronchetti83,Tanaka12}, locally favored structures (LFS) \cite{Tanaka08, Royall13, Royall13_2, Williams15, Royall15, Royall15_2, Royall16, Royall18, Tanaka13}, excess entropy \cite{Parrinello17_1, Parrinello17_2, Arratia20, Tanaka18}, machine-learning-based measures of local softness \cite{Liu15, Liu16, Liu17, Reichman20, Liu23}, and, more recently, mean-field caging potentials derived from the arrangement of nearest neighbors \cite{Bhattacharyya17, Bhattacharyya21, Bhattacharyya22}.\\

A key concept to emerge from early investigations of supercooled liquids and metallic glasses is polytetrahedral ordering. In a seminal work, Frank \cite{Frank52} pointed out that, for a cluster of 13 atoms in a Lennard-Jones system, the icosahedral arrangement is the locally densest packing, exhibiting lower free energy than crystalline nuclei with FCC or HCP structures. These icosahedral motifs are energetically favorable and long-lived but cannot tile space without introducing defects. This geometric incompatibility gives rise to frustration, which inhibits crystallization. These insights laid the foundation for understanding the structure of complex alloys and quasicrystals \cite{Kasper63} and inspired a theoretical framework based on three-dimensional disclinations, which explained how polytetrahedra are incompatible with periodic order and provided deeper insights into glass formation \cite{Nelson83}. Subsequent studies confirmed the presence of such locally favored structures in bulk liquids and emphasized the role of icosahedral ordering in the emergence of slow dynamics in glassy systems \cite{Tarjus03, Ronchetti81, Ronchetti83, Andersen88, Tsumuraya91, Tarjus12, Chen13, Ma14, Smallenburg19, Coslovich20}. \\

A notable development in the identification of structural motifs was the introduction of the topological cluster classification (TCC) algorithm \cite{Royall13, Royall13_2}, designed to detect polytetrahedral structures resembling low-energy configurations in systems with a variety of interaction potentials. This method has been instrumental in elucidating the role of LFS in dynamical arrest in gels \cite{Tanaka08, Tanaka08-2, Tanaka15} and supercooled liquids \cite{Royall15, Royall18, Kob17}. Other studies have proposed simpler tetrahedra as fundamental structural units and have defined order parameters based on particle tetrahedrality \cite{Pastore07, Smallenburg20, Wang15, Wang18, Foffi22, Tanaka18-1, Medvedev07, Filion20}. Particle-resolved studies of colloidal gels and supercooled liquids have enabled direct experimental investigation of LFS, providing valuable insight into the interplay between structure and dynamics \cite{Tanaka08, Royall18}. The analysis of connected networks of polytetrahedra in dense colloidal suspensions using the TCC method has revealed a growing static length scale that evolves in tandem with a dynamic length scale at the onset of glassy dynamics \cite{Royall18}. This approach has also uncovered the polytetrahedral structural organization in colloidal gels and has offered insights into their mechanical failure \cite{Jack23}. \\

In recent years, topological aspects of plasticity in amorphous solids, particularly those linked to polytetrahedral structures, have received increasing attention \cite{Ma14_2, Royall16_2, Royall18_2}. For instance, a recent numerical study of glass-forming liquids under linear shear revealed strong correlations between plastic deformation and regions deficient in LFS \cite{Jack23}. Similar conclusions have emerged from simulations of glasses under cyclic shear deformation, using two-body excess entropy and local tetrahedrality as structural indicators \cite{Foffi22}. However, experimental studies in this direction remain scarce, with the notable exception of sheared granular glasses \cite{Wang18}. This study considered tetrahedra as building blocks and demonstrated that plastic deformation arises from topological changes in these structures.\\
 
In a parallel development, recent investigations have focused on identifying locally soft/weak regions, which are susceptible to plastic rearrangements, based on structural measures. The machine learning approaches have shown some success in this direction \cite{Liu15,Liu16,Liu17,Liu23}, while some of the present authors identified weak/soft zones using a structural order parameter microscopically derived from the mean-field local caging potential experienced by a particle due to its neighbors \cite{Bhattacharyya21,Bhattacharyya22,Chikkadi24}. The structurally stable regions exhibited deeper caging potential compared to unstable regions. These observations raise the question of whether mean-field caging potentials can provide further insight into the stability of polytetrahedral motifs in dense colloidal suspensions. Are icosahedrally ordered neighborhoods associated with deeper caging potentials than those in crystalline or other structural motifs? Furthermore, experimental investigations into the response of locally favored structures to applied shear remain limited. Closing this gap would bridge our understanding of dynamic relaxation via a microscopically motivated structural order parameter to local particle topology, thus enhancing our understanding of how amorphous solids resist deformation and ultimately fail under load.\\

\section*{Results}
\begin{figure*}[t!]
\centering
\begin{tabular}{lcr}
    \includegraphics[width=0.35\textwidth]{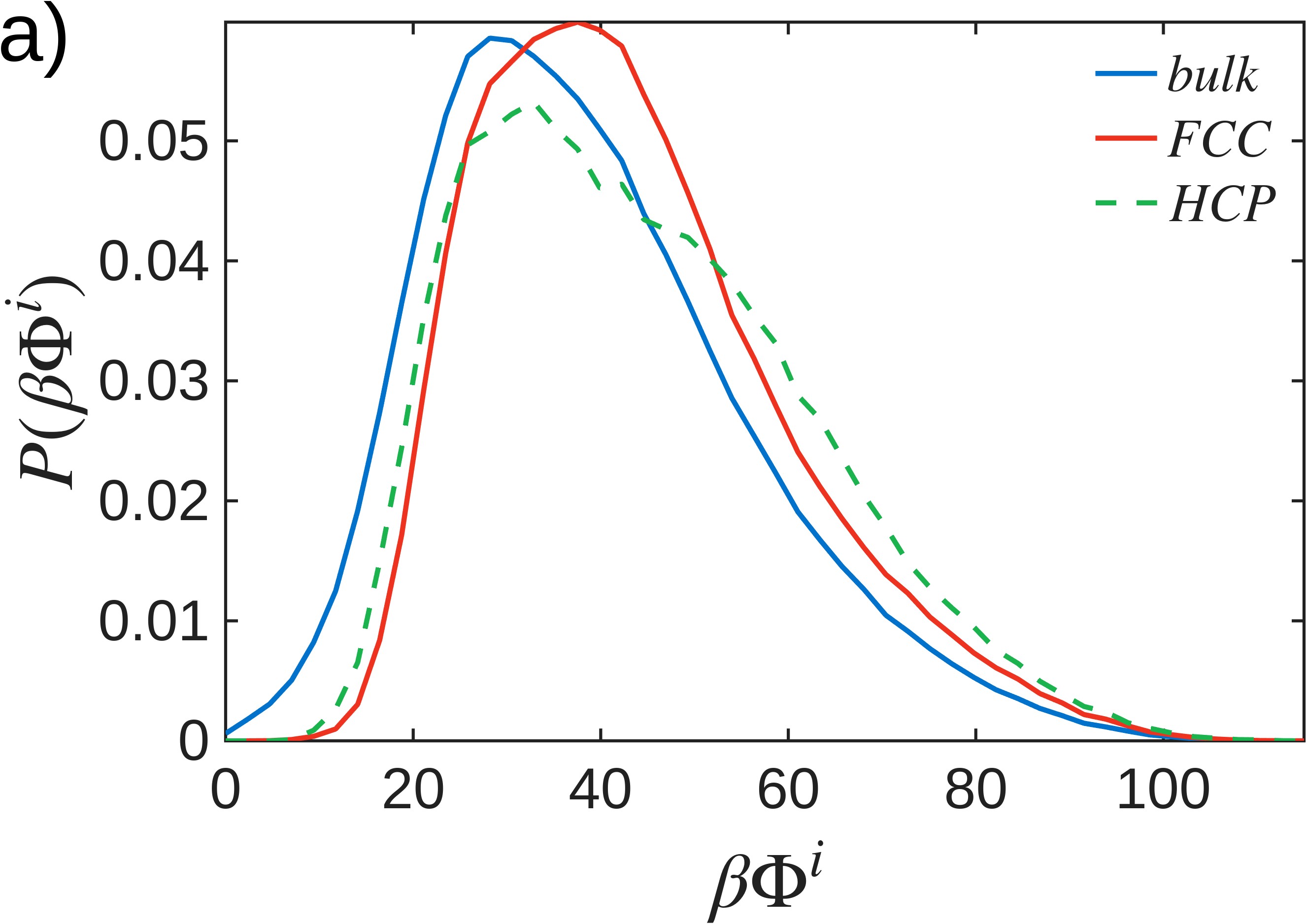} &
    \includegraphics[width=0.35\textwidth]{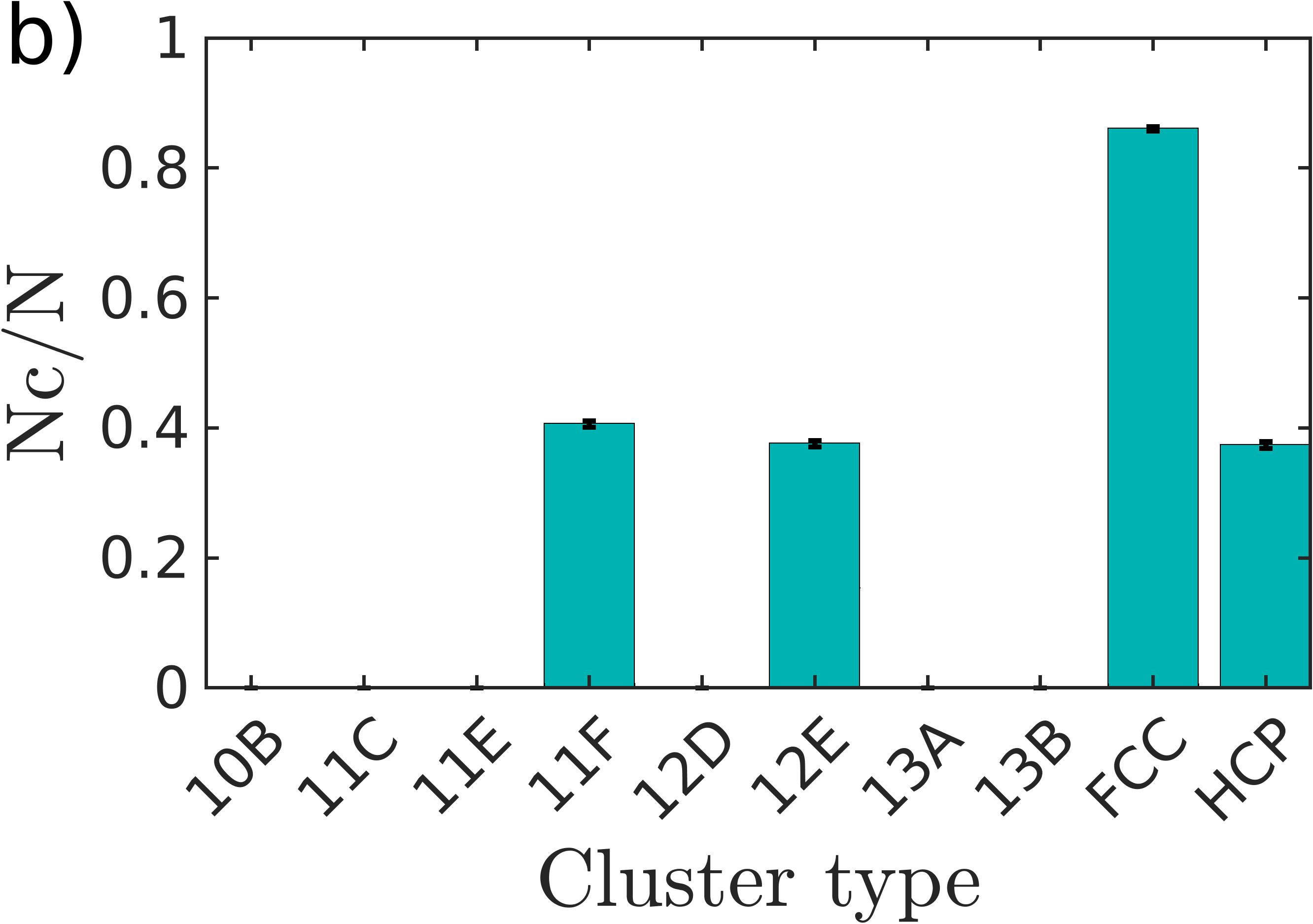} &
    \includegraphics[width=0.26\textwidth]{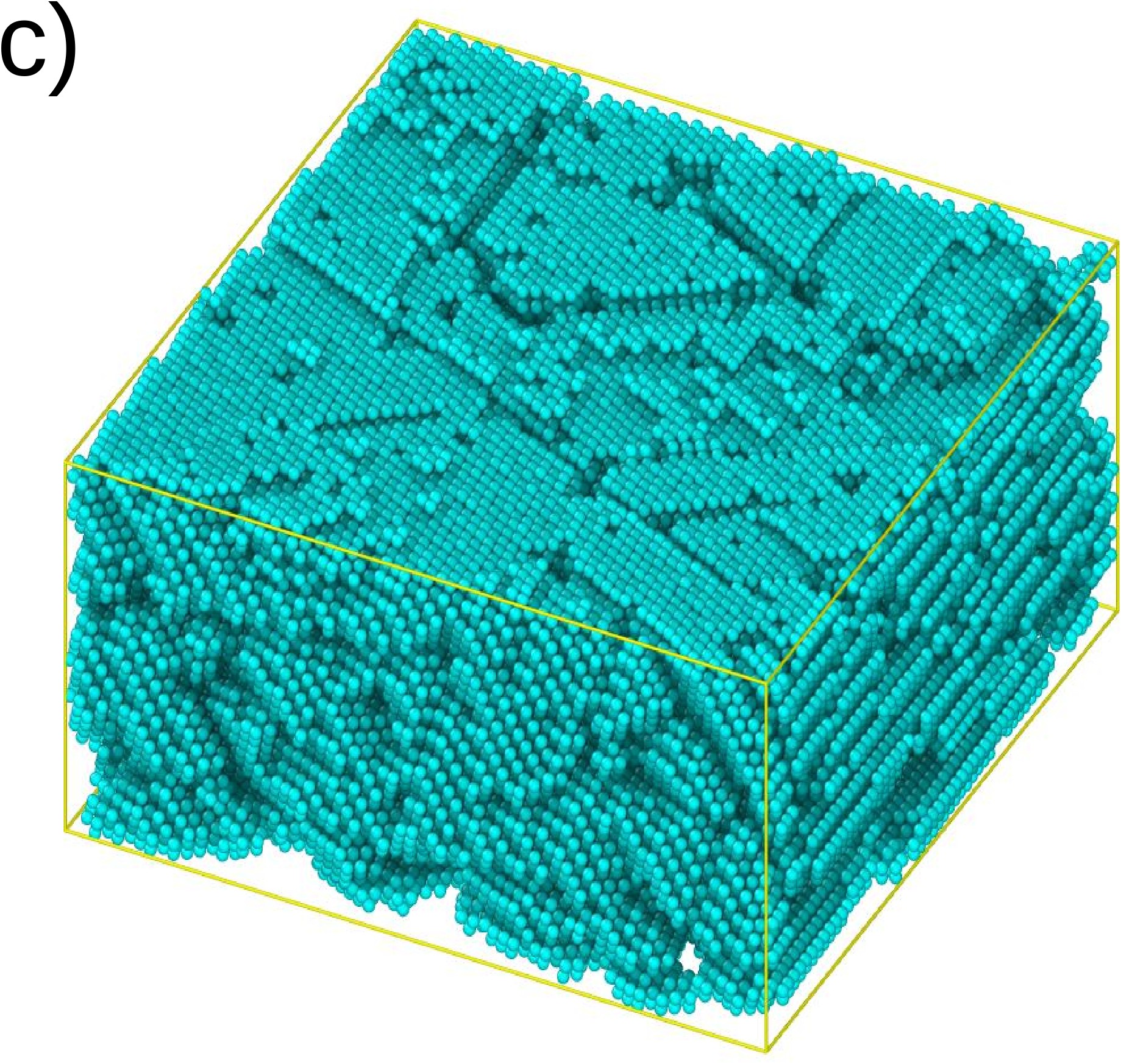} 
\end{tabular}
\caption{Topological structures and caging potential of stable motifs found in colloidal crystals. (a) Caging potential of the bulk particles compared to the particles in different motifs. The blue curve represents all particles, red for FCC and dashed green for HCP respectively. (b) Fraction of particles found in different motifs, with error bars showing standard deviations. (c) Particles found in face-centered cubic structure from the topological analysis. }
\label{Fig1}
\end{figure*}

The colloidal crystals consist of monodisperse silica beads with a diameter of $1.5,\mu\mathrm{m}$ at a volume fraction of $\phi \sim 0.53$. Details of the preparation protocol are provided in the Methods section. Colloidal glasses were prepared using sterically stabilized, fluorescent polymethylmethacrylate (PMMA) particles with a diameter of $\sigma = 1.3,\mu\mathrm{m}$ and a polydispersity of 7\%, which effectively suppresses crystallization. The suspensions, imaged using confocal microscopy, had a volume fraction of $\phi \approx 0.60$ and contained approximately 100,000 particles. Further experimental details are included in the Methods section. To identify relevant structural motifs, we employed the Topological Cluster Classification (TCC) algorithm, as detailed in the Methods section. A library of motifs identified using this approach is presented in Fig.~S1 of the Supplementary Information.\\

We begin by characterizing the topological features of stable motifs in quiescent colloidal crystals and glasses, quantifying their stability via the mean-field local caging potential. Following this, we extend the analysis to sheared colloidal glasses to examine topological changes and the role of caging potentials, thereby uncovering the structural fingerprints of relaxation under deformation.

\subsection*{Locally favored structures and their caging potential in colloidal crystals}
We compute the mean-field caging potential to analyze the relative stability of the structural motifs found in colloidal crystals. Since our experiments do not provide direct access to the system's free energy, we infer stability from the caging potential. Deeper is the caging potential, stabler is the motif. The caging potential $\Phi^i$ of a colloidal particle, based on the arrangement of its neighbors \cite{Bhattacharyya22, Chikkadi24}, is given by:
\begin{equation}\label{eq:1}
\beta \Phi^{i}= \rho \int \mathbf{dr}~C^{i}(r) g^{i}(r),
\end{equation}
\noindent
where $\rho$ is the density, $g^{i}(r)$ is the particle-level radial distribution function (RDF), and $C^{i}(r)$ is the direct correlation function, which is approximated as $C^{i}(r) \approx g^i(r) - 1$ (see Methods for details). The depth of the caging potential in Eq.~\eqref{eq:1} is obtained from microscopic density functional theory \cite{Bhattacharyya21, Bhattacharyya22}. It admits a simple and intuitive interpretation: the RDF encodes information about the local configuration of particles, while the direct correlation function represents the effective short-range interaction between a particle and its neighbors. The product of these two functions yields the caging potential experienced by a particle due to its surroundings. More structured neighborhoods are associated with deeper caging potentials, indicating greater local stability, while disordered neighborhoods correspond to shallower potentials. We use this structural metric to assess the relative stability of various structural motifs. The distribution of caging potentials for different crystalline motifs is shown in Fig.~\ref{Fig1}a. The blue curve represents the distribution for all particles in the system, referred to as the bulk, the red curve corresponds to the FCC structure, and the dashed green curve represents the HCP structure. Both FCC and HCP exhibit higher caging potentials than the bulk, indicating enhanced local stability. The FCC motif is the most dominant, accounting for nearly $86\%$ of the particles \cite{Royall13_2}. This is shown in Fig.~\ref{Fig1}b, which displays the fraction of particles in each of the motifs along with black vertical lines on the bars representing the standard deviation calculated using multiple configurations of the system. A reconstruction of the particles that are part of FCC motif is shown in Fig.\ref{Fig1}c. In addition, we find other motifs such as HCP, $12E$ and $11F$ present in similar fractions. The $12E$ and $11F$ motifs are distorted forms of the crystalline motifs with one or more particles missing from the from the six-member rings, combined with distortions of the topology. We refer to these motifs as defective forms of regular crystal motifs, see Fig.~S1 in Supplementary Information for details.

\subsection*{Locally favored structures and their caging potential in colloidal glasses}

\begin{figure*}[t!]
\centering
\begin{tabular}{cc}
\includegraphics[width=.38\textwidth]{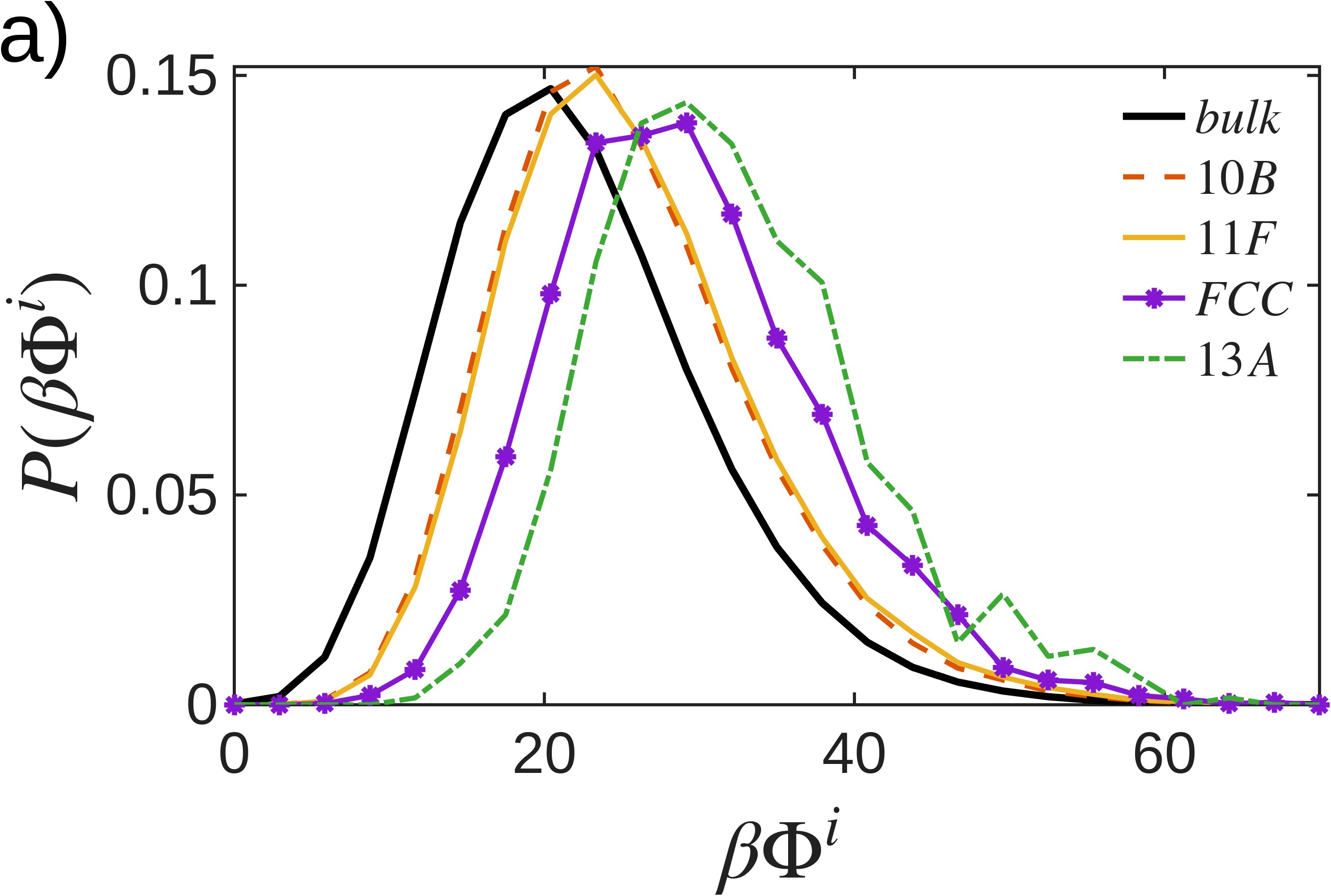}&
\includegraphics[width=.25\textwidth]{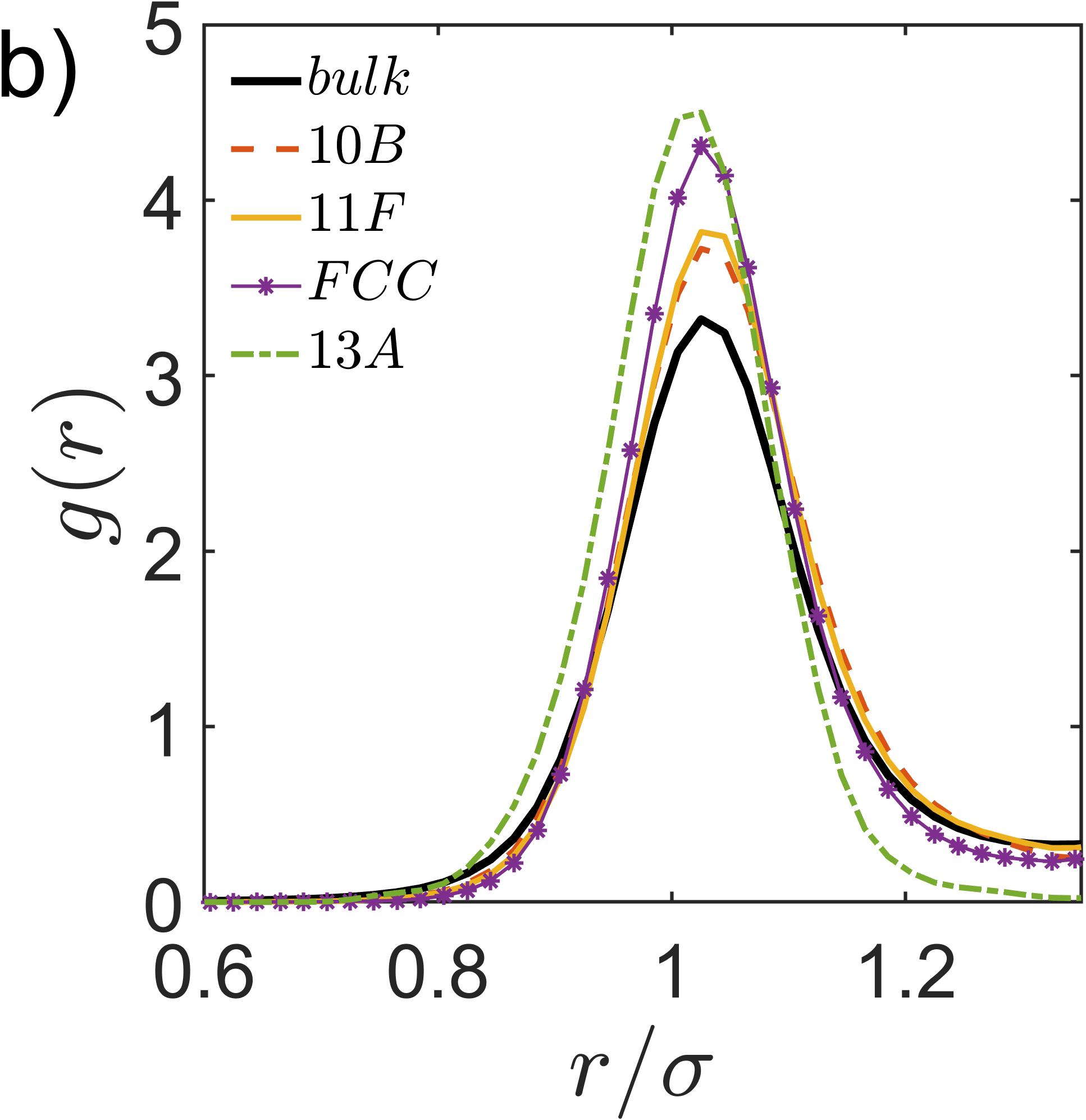}\\
\includegraphics[width=.36\textwidth]{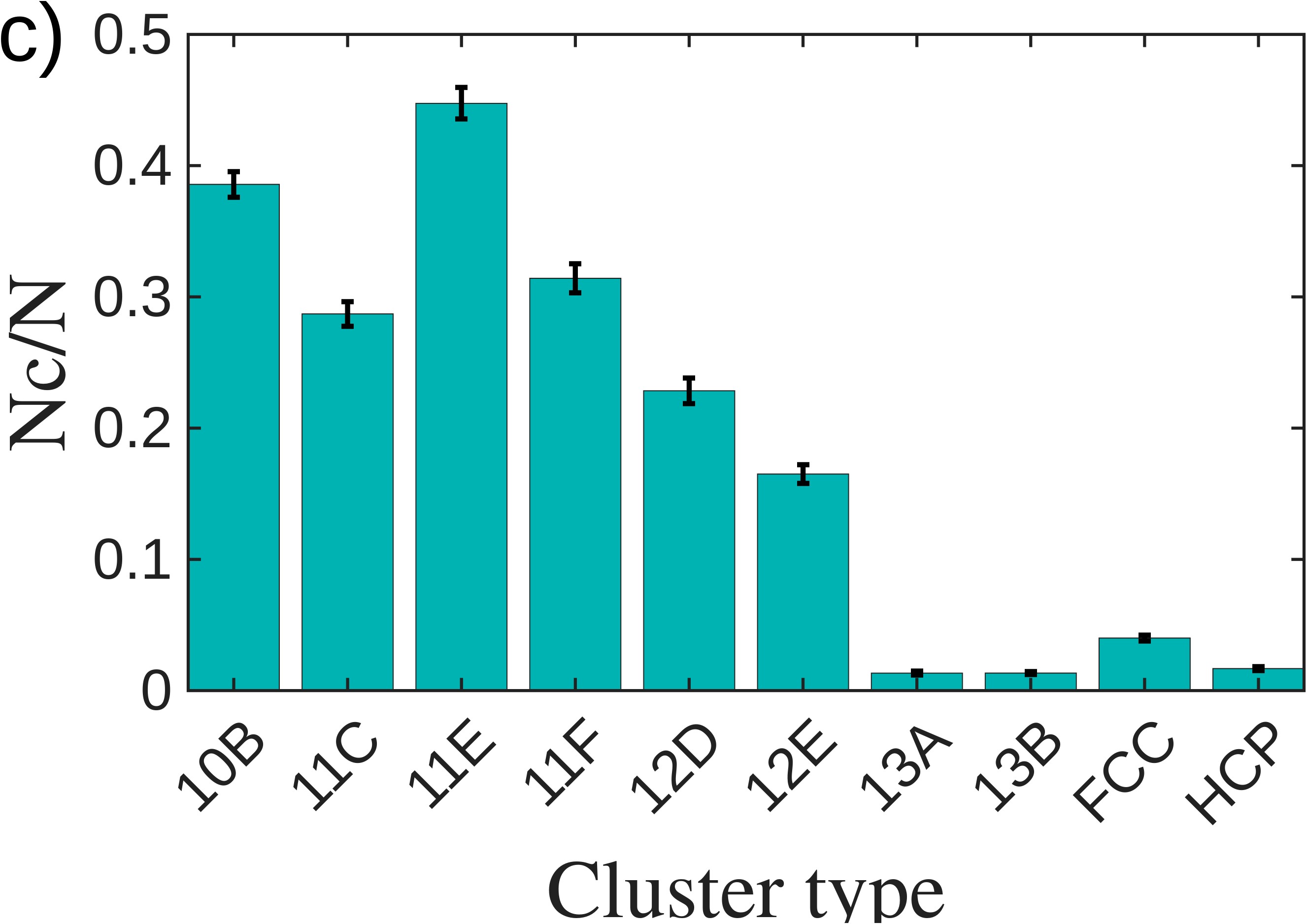}&
\includegraphics[width=.32\textwidth]{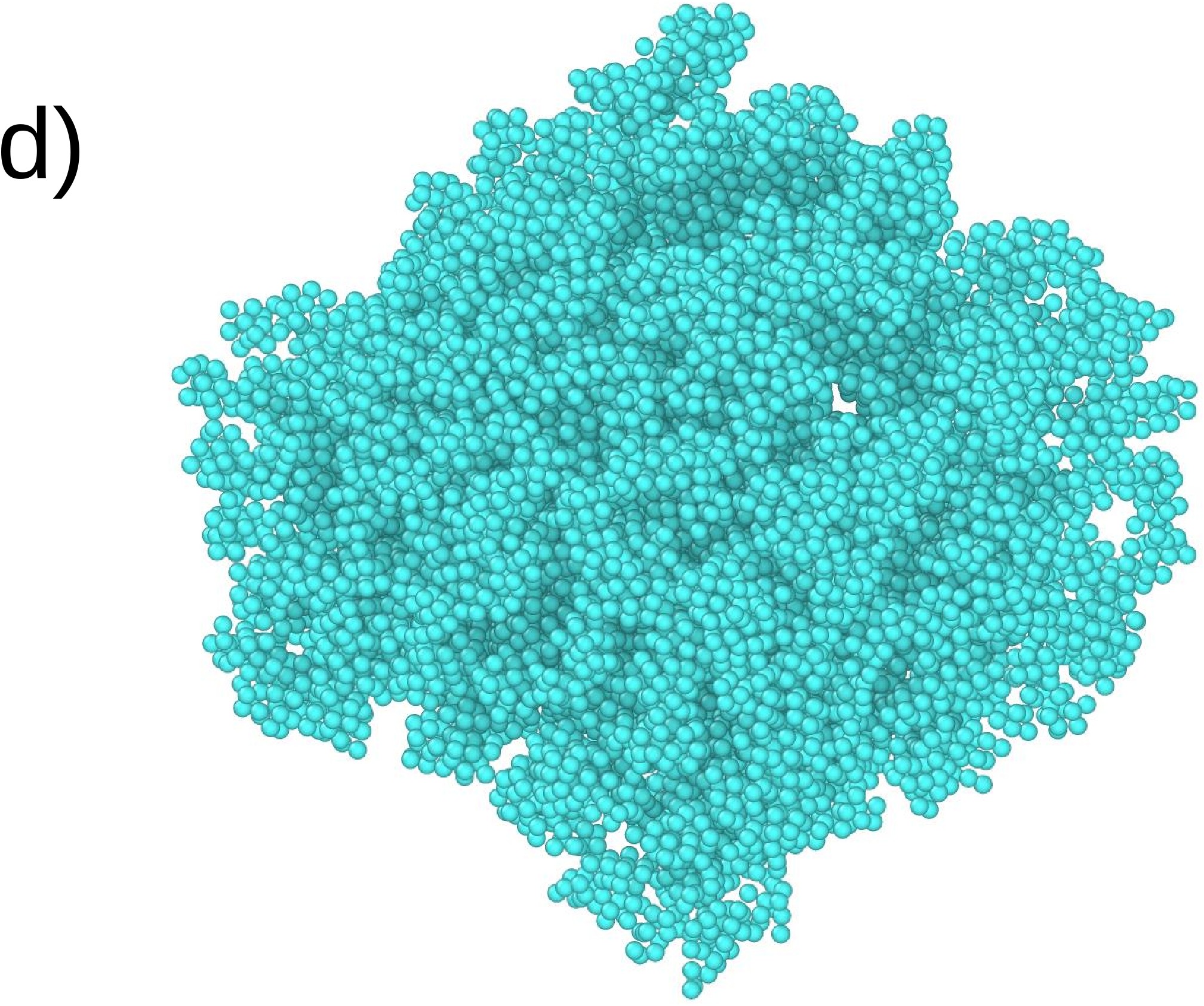}\\
\end{tabular}
\caption{(a) Caging potential of different clusters compared to that of the bulk system. The distributions are averaged over $10$ configurations. (b) Pair correlation function of particles belonging to different cluster types. (c) Fraction of particles found in different cluster types and the error bars are the standard deviations in the values. (d) Particles belonging to $10B$ structure in a dense suspension of PMMA colloidal particles at a packing fraction of $\phi=0.60$. Note that only the central particles were considered in calculation of c) and d).}
\label{Fig2}
\end{figure*}

Colloidal glasses display a variety of motifs containing both five- and six-membered ring structures. Regular motifs in the packing include fivefold symmetric icosahedral clusters of thirteen particles, referred to as $13A$, as well as crystalline motifs of $FCC$ and $HCP$ structures. To compare their relative stabilities, we analyze the distribution of the mean-field caging potential, $P(\beta\Phi_i)$, shown in Fig.~\ref{Fig2}a, see Methods for the details. The thick black line corresponds to the distribution of all particles in the system or the bulk, the dotted line corresponds to icosahedral structures and the line with asterisk symbols represented the FCC structure. Interestingly, the icosahedral motif $13A$ is associated with the deepest caging potential, larger than that of the $FCC$ structure, therefore making it the most stable motif. In general, the motifs with lower energies are expected to be associated with higher local density. To examine this aspect, we computed the pair correlation function $g(r)$. First, $g(r)$ was calculated for all particles or the bulk. Then, for each cluster type, only the central particles were tagged to obtain the pair correlation function. The results in Fig.~\ref{Fig2}b indeed show that particles in icosahedral clusters have more densely packed neighborhoods compared to those in $FCC$ clusters or in the bulk. Together, these results reaffirm that icosahedral motifs are associated with lower energy and greater stability than crystalline motifs.\\

Although particles in $13A$ and $FCC$ motifs exhibit deeper caging potentials and higher mechanical stability, their populations are relatively small (see Fig.~S2a and Fig.~S2b in the Supplementary Information for cluster snapshots). Instead, the amorphous phase is dominated by defective five-membered ring structures such as $10B$, $11C$, $11E$, and $12D$, along with defective crystalline structures like $11F$ and $12E$ (see Fig.~S1 for a summary of motifs identified by the TCC algorithm \cite{Royall13}). The fractions of particles in different motifs are displayed in Fig.~\ref{Fig2}c. Fewer than $2\%$ of particles form regular icosahedra ($13A$), while crystalline motifs ($FCC$ and $HCP$) together account for less than $10\%$ of the system. The most dominant motifs are $11E$ and $10B$, followed by $11F$ and $11C$. Thus, the amorphous system primarily consists of defective five-membered ring structures and defective crystalline motifs. The distributions of caging potential for the dominant defective motifs ($10B$ and $11E$) are included in Fig.~\ref{Fig2}a. These motifs exhibit deeper caging potentials than bulk particles, but shallower than those of the regular motifs $13A$ and $FCC$. Similar trends are evident in the height of the first peak of $g(r)$ in Fig.~\ref{Fig2}b, where larger values compared to the bulk indicate that these motifs are stable structures that contribute to the rigidity of the amorphous phase. This is further supported by instantaneous snapshots of $10B$ clusters in Fig.~\ref{Fig2}d, which reveal a connected network spanning across the field of view. Earlier simulations of polydisperse hard-sphere systems identified similar motifs \cite{Royall15_2, Royall15}, where particles with $10B$ structure was found to be long-lived and spanning the system. Subsequent experimental studies on supercooled colloidal liquids further emphasized the structural and dynamical role of $10B$ motifs \cite{Royall18}. Motivated by these findings, we focus our analysis on clusters of $10B$ particles—abundant five-membered ring structures—and the $11F$ motif, which is a defective crystalline structure.

\begin{figure*}[t!]
\centering
\begin{tabular}{cc}
\includegraphics[width=.36\textwidth]{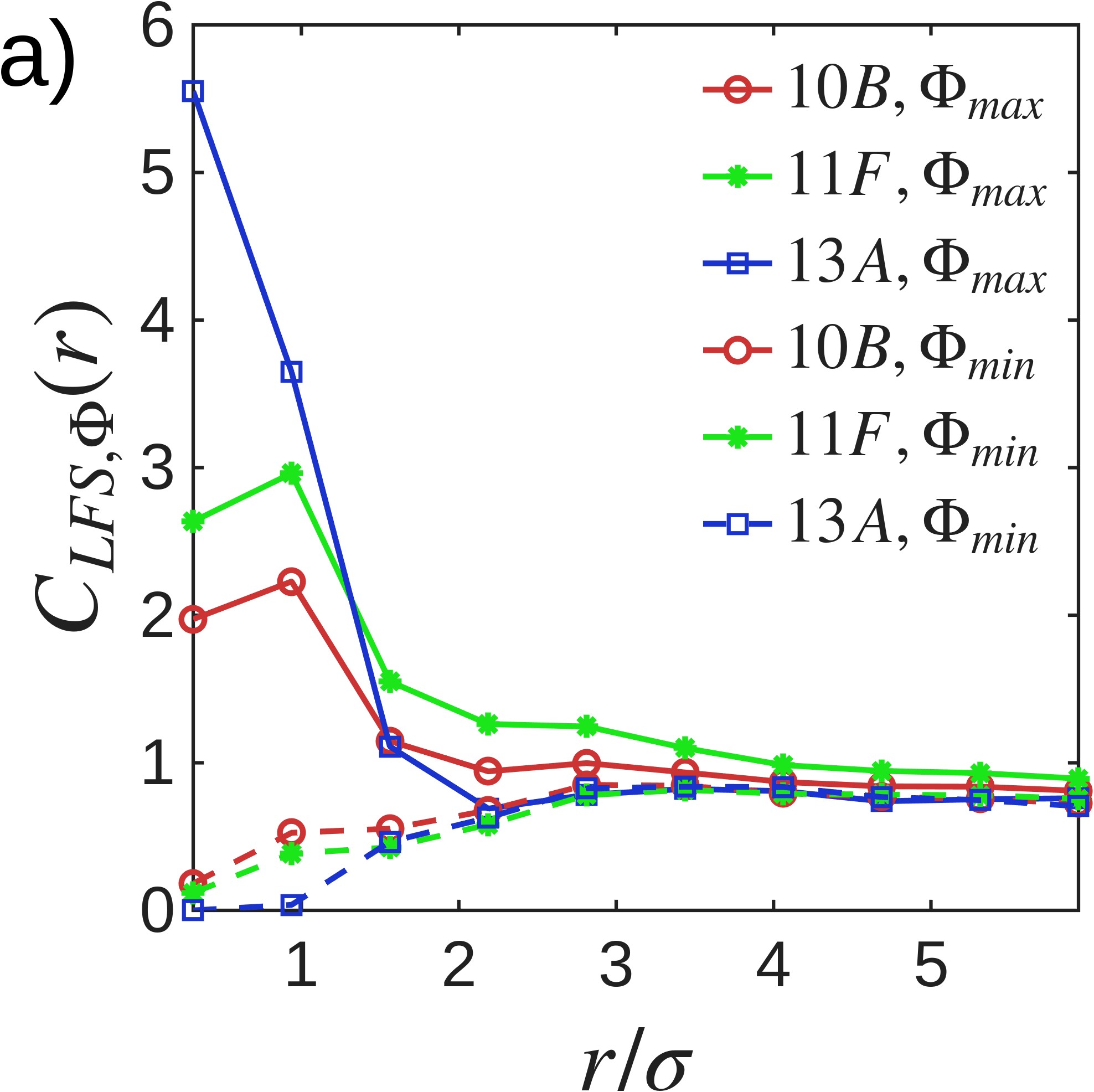}&
\includegraphics[width=.43\textwidth]{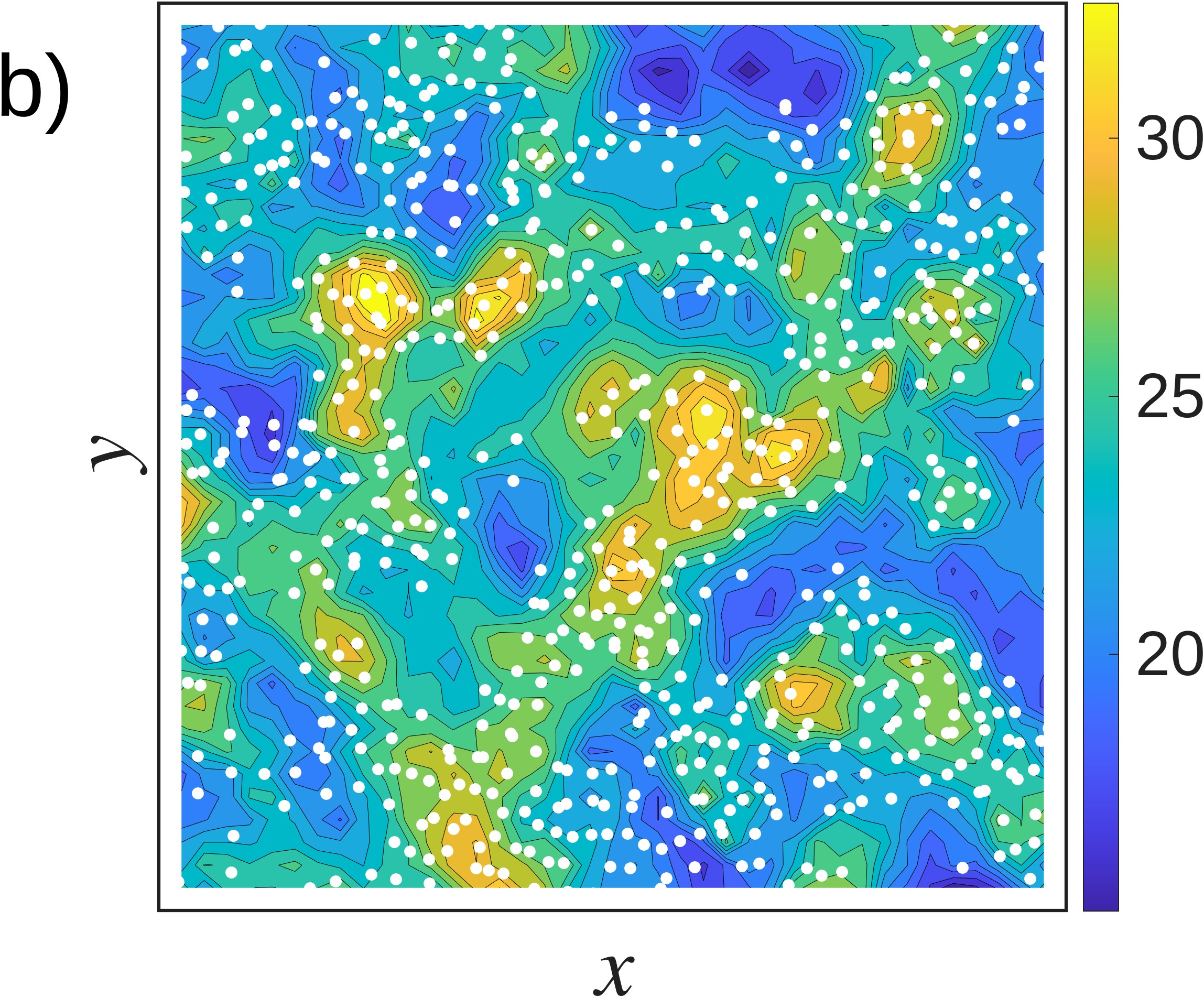}
\end{tabular}
\caption{(a) The spatial correlation of the LFS particles with soft (lowest $10\%$ caging potential) and hard zones (highest $10\%$ caging potential) is shown with dashed and solid lines, respectively. Red circles represent $10B$ particles, green stars represent $11F$ particles, and blue squares represent $13A$ particles. (b) A coarse-grained map of the caging potential obtained using all particles in a cross-section of thickness $2\sigma$, with $10B$ particles overlaid as white circles. The caging potential is coarsegrained over the first nearest neighbors.}
\label{Fig3}
\end{figure*}

For further understanding of the heterogeneous structure of colloidal glasses, we correlate the caging potential (CP), which identifies locally stiff (high $\Phi$) or soft environments (low $\Phi$), directly with the abundance of characteristic structural motifs. These correlations are  quantified using the radial pair correlations between locally favored structures (LFS) and particles with deeper caging potential (stable zones) as well as those with lower caging potential (unstable zones), see Methods section for details. The correlation function, $C_{LFS,\Phi}(r)$, between particles in $10B$, $11F$ and $13A$ motifs and particles with $10\%$ highest caging potential is shown as solid lines, while the correlation with $10\%$ lowest caging potential particles is represented by dashed lines in Fig.~\ref{Fig3}a. The presence of a peak at small $r$ in $C_{LFS,\Phi}(r)$ between hard particles and LFS density demonstrates a strong spatial correlation between the two quantities. These results clearly show that particles belonging to stabler motifs indeed belong to hard zones with deeper caging potentials. A broader comparison across different structural motifs provided in the supplementary figure Fig.~S5 point to similar conclusions. A visual impression of these correlations is presented in in Fig.~\ref{Fig3}b, where a coarse-grained map of the caging potential obtained using particles in a $2\sigma$ thick section is displayed with the positions of $10B$ particles overlaid as white circles. The yellow regions correspond to hard zones with deeper caging potential. We observe that clusters of $10B$ particles preferentially fall in these zones. Similar maps for other motifs are shown in Fig.~S4 of the Supplementary Information. \\

\section*{Topological changes of ordered structures due to applied shear}

\begin{figure*}[t!]
\centering

\begin{minipage}{0.5\textwidth}
    \centering
    \includegraphics[width=\linewidth]{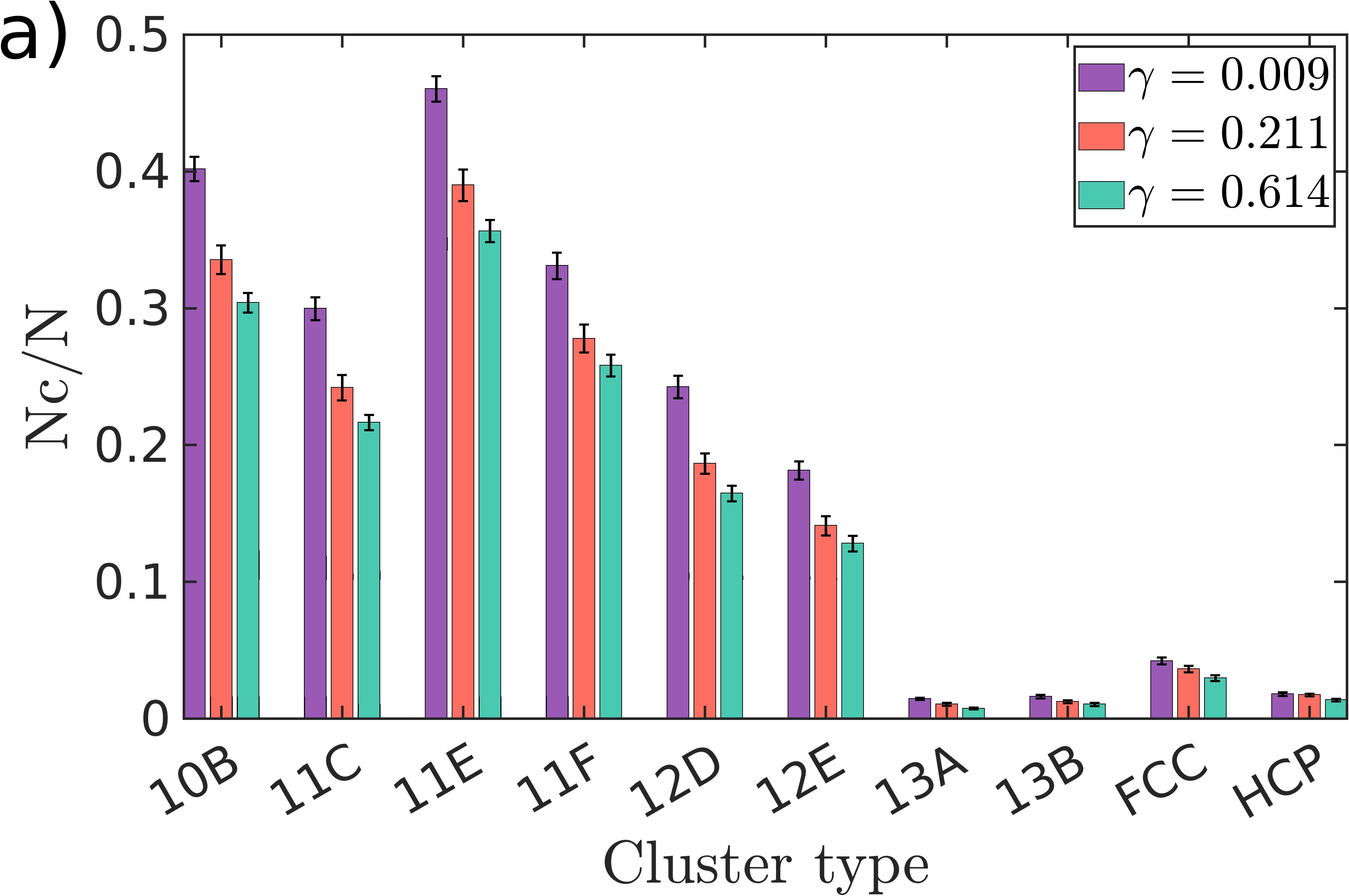}
\end{minipage}
\hspace{0.05\textwidth}
\begin{minipage}{0.35\textwidth}
    \centering
    \includegraphics[width=\linewidth]{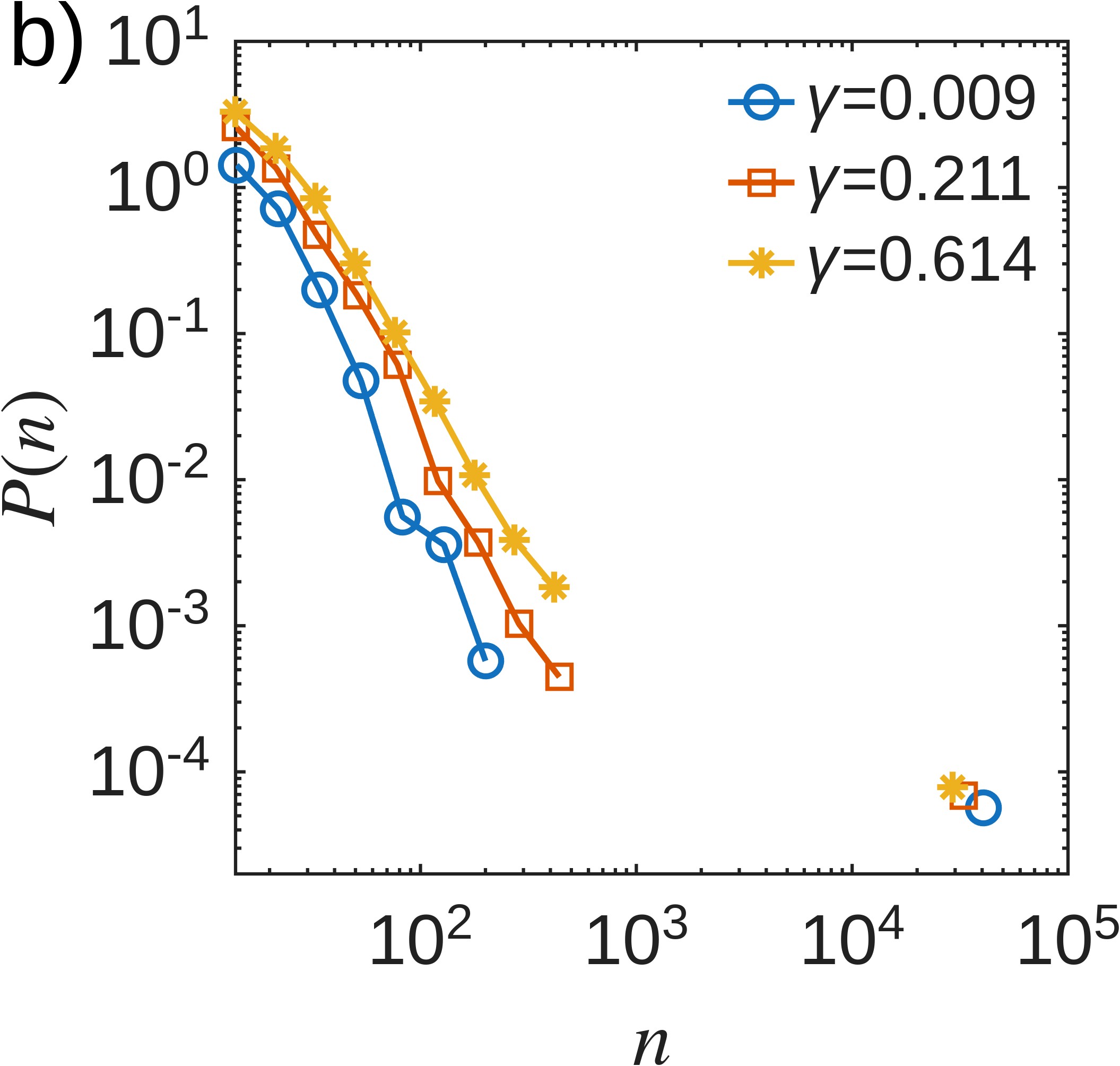}
\end{minipage}

\vspace{1em}

\begin{minipage}{0.33\textwidth}
    \centering
    \includegraphics[width=\linewidth]{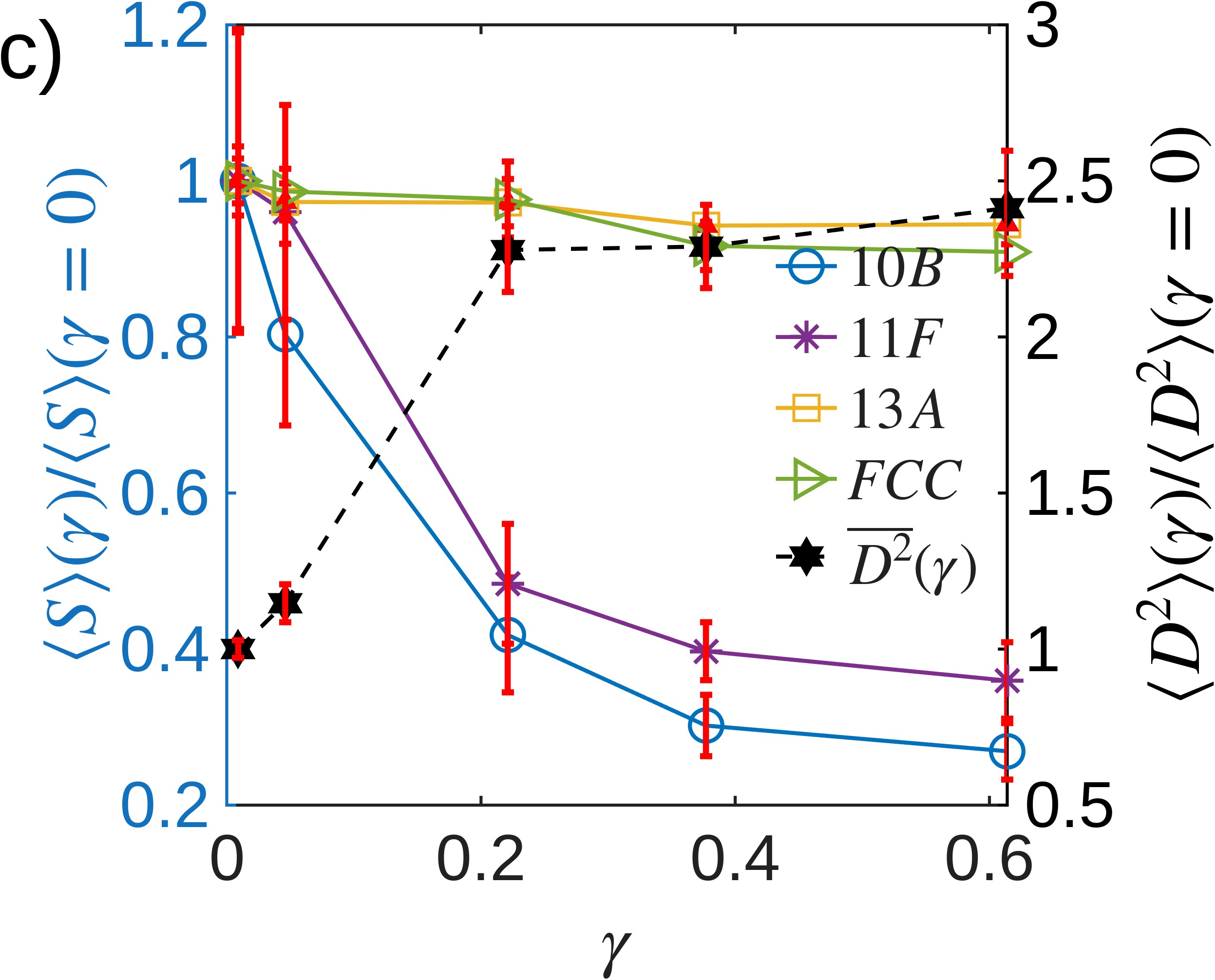}
\end{minipage}
\hspace{0.02\textwidth}
\begin{minipage}{0.3\textwidth}
    \centering
    \includegraphics[width=\linewidth]{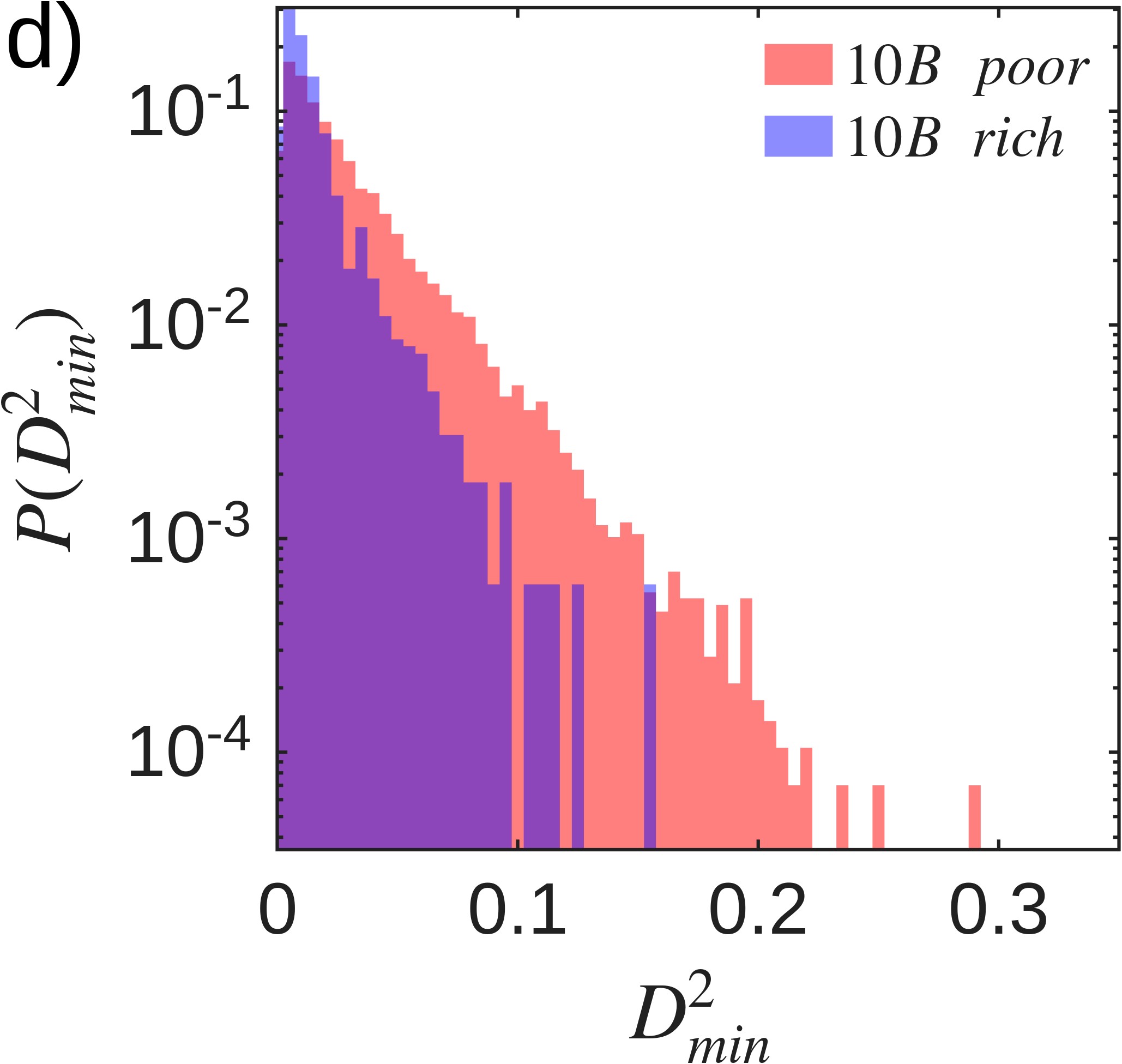}
\end{minipage}
\hspace{0.02\textwidth}
\begin{minipage}{0.3\textwidth}
    \centering
    \includegraphics[width=\linewidth]{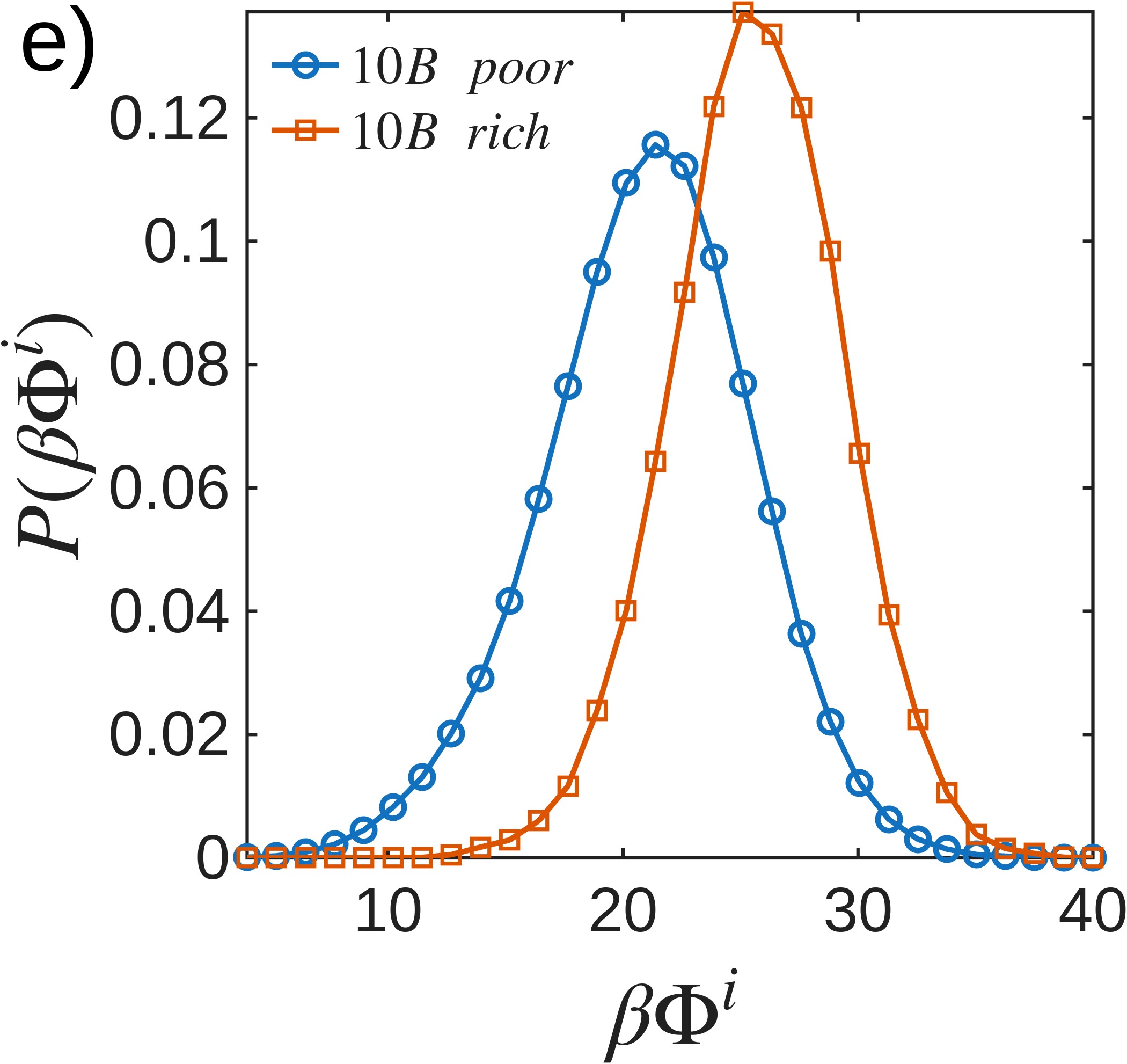}
\end{minipage}

\caption{(a)Fraction of particles found in different cluster types as a function of strain. The height represents the mean and the black vertical error bar is the standard deviation after averaging over a small strain window of $\Delta \gamma=0.018$. (b) cluster size distribution of the $10B$ particles for different strain. (c) Scaled mean cluster size of the particles as the system is sheared is on the left-y axis for $10B$(blue circle) $11F$(purple star) $13A$ (yellow square) and FCC (green triangle). The  scaled mean non-affine displacement of the system is plotted on the right-y axis in black dashed line. The vertical red error bar presents the standard deviation to the mean values in several snapshots in the strain window $\Delta \gamma=0.018$. (d) Distribution of non-affine displacement for the $10B$ rich and $10B$ poor trajectories. (e) Histogram of caging potential for the $10B$ rich and $10B$ poor trajectories. Both d) and e) ware calculated in the steady state at $\gamma=0.614$}
\label{Fig4}
\end{figure*}


We now seek to understand how the structural motifs in the system respond to external shear. First, we analyze the variation of the fraction of particles, $N_c/N$, in different motifs as a function of strain. As illustrated in Fig.~\ref{Fig4}a, this fraction decreases with increasing strain, indicating that particles gradually lose their structural order under shear. Due to abundance of defective motifs, the particles associated with these structures naturally exhibit a pronounced response to applied shear. To investigate this further, we examine the changes in the cluster size distribution (CSD), $P(n)$, under shear. Fig.~\ref{Fig4}b shows the CSD of particles belonging to the $10B$ motif at different strain values (see Methods for details on cluster identification). The presence of an isolated point at large-$n$ in the tails of the distributions is another evidence of system spanning network of particles in defective motifs. 
Upon shearing, particles tend to leave the largest cluster and either join smaller clusters or lose structural order altogether. This feature is evident in the evolving CSD: the size of the largest cluster decreases, while the number of smaller clusters increases with increasing strain. A similar trend is observed for other motifs as well, see Fig.~S6 in supplementary information. We next present the average cluster size for selected motifs as a function of strain in left-y axis of Fig.~\ref{Fig4}c, which includes the contribution from the largest cluster. Clearly, the scaled average cluster size, $\left<S(\gamma)\right> / S(\gamma = 0)$, for both $10B$ and $11F$ motifs decreases with strain and saturates beyond $\gamma = 0.2$. The Fig.~\ref{Fig4}c also displays the non-affine displacements, $D^2_{\min}$, of particles on the right-y axis, which serve as a dynamic measure of plasticity (see Methods for details). Strikingly, both the structural measure (average cluster size) and the dynamical measure ($D^2_{\min}$) saturate beyond a strain of $\gamma = 0.2$, suggesting structure-dynamics correlations. A similar correlation between the inverse mean-field caging potential (structural) and $D^2_{\min}$ was reported in an earlier study \cite{Chikkadi24}.\\


Next, we investigate in detail the topological changes induced by shear by analyzing the non-affine displacements of particles. While non-affine displacements typically signal plastic deformation, they are also accompanied by structural rearrangements. To test whether particles that lose topological order under shear are associated with large non-affine displacements, we classify particle trajectories into two categories: LFS-rich and LFS-poor. A particle is considered to follow an LFS-poor trajectory if it belongs to a locally favored structure in at least one configuration but in fewer than $25\%$ of the recorded configurations during shear. In contrast, a particle that remains part of a stable motif throughout its entire trajectory ($100\%$ of the time) is categorized as LFS-rich. Figure~\ref{Fig4}d shows the distribution of non-affine displacements, $P(D^2_{\min})$, for particles associated with the $10B$ motif. Strikingly, particles on $10B$-poor trajectories exhibit significantly larger non-affine displacements compared to those on $10B$-rich trajectories. In our recent work \cite{Chikkadi24}, we demonstrated that particles undergoing plastic rearrangements or large non-affine displacements tend to reside in regions with shallower local caging potentials, $\Phi^i$. Building on this result, we now examine the distribution of local caging potentials, $P(\beta\Phi^i)$, for particles on $10B$-rich and $10B$-poor trajectories. As shown in Fig.~\ref{Fig4}e, particles that lose structural order under shear are consistently associated with lower caging potentials. Importantly, these findings are robust across different structural motifs (see Supplementary Information).

Previous studies on sheared granular glasses had linked structural instability to plasticity by analyzing the shape of tetrahedra \cite{Wang18}. In contrast, our study identifies unstable motifs through the local caging potential of constituent particles. A key advantage of our structural metric, the mean-field caging potential ($\Phi^i$), lies in its broad applicability to both crystalline and amorphous systems. Together, these findings establish that the loss of mechanical stability under shear is intimately connected to topological transformations that facilitate the breakdown of structural order, thereby enabling plastic deformation.\\

\section*{Conclusions}
In summary, we have established clear correlations between the stability of topological motifs and the local caging potential in quiescent colloidal crystals and glasses, as well as in weakly sheared colloidal glasses. Our analysis reveals that colloidal crystals formed in our experiments predominantly exhibit FCC ordering, along with the presence of HCP structures and other defective motifs. Notably, no five-membered ring structures are observed. In contrast, colloidal glasses display an abundance of defective icosahedral motifs and defective crystalline structures. An analysis of their stability, based on the local caging potential, reveals that regular icosahedral clusters ($13A$) are associated with the deepest caging potentials—lower than those of crystalline motifs such as FCC and HCP. Consequently, particles in locally favored structures are preferentially located in regions with deeper caging potential, indicating strong spatial correlations between local structure and mechanical stability.\\

Furthermore, we demonstrate that shear-induced loss of mechanical stability in colloidal glasses is accompanied by significant topological changes in the clusters of stable motifs. The fraction of particles residing in such locally favored structures decreases progressively with increasing strain. This results from the particles leaving the larger clusters and losing their structural order. We find that particles that become disordered under shear exhibit larger non-affine displacements and shallower caging potentials, making them more susceptible to plastic rearrangements.\\

These findings underscore the critical role of local structural order in governing the mechanical stability of disordered solids. They provide a framework for understanding how amorphous materials resist deformation and maintain rigidity. The insights obtained from this study could be valuable for rational design of amorphous materials with tailored mechanical properties.\\

\section*{Materials and Methods}

\subsection{Experimental realization of 3D colloidal crystals} 
The 3D crystal was created by suspending silica beads of $\sigma= 1.5 \mu m$ in a $64:36$ mixture of dimethyl sulfoxide and deionized water, to match the refractive index of the particles. To enable 3D imaging, an aqueous suspension of fluorescein salt is added to the solvent. Due to a density mismatch, the particles sediment under gravity, and after $\sim 20h$ hours of equilibration, the system spontaneously forms a face-centered cubic (FCC) crystal structure. The imaging was done using Leica-Dmi8 confocal microscope with a field of view of $92*92*45 \mu m$ containing nearly $100000$ particles. The packing fraction of this system is $\phi \sim 0.53$. \\

\subsection{Experimental techniques used for performing shear measurements} 
The shear experiments are performed using dense colloidal suspensions of sterically stabilized fluorescent polymethylmethacrylate particles in a density and refractive index matching mixture of cycloheptyl bromide and cis-decalin. The particles have a diameter, $\sigma = 1.3 \mu m$ and a polydisperity of $7\%$ to prevent crystallization. The suspension was centrifuged at an elevated temperature to obtain a dense sediment, which was subsequently diluted to get a suspension of the desired volume fraction $\phi\sim0.60$. The sample was sheared using a shear cell with two parallel boundaries separated by a distance of $\sim 50\sigma$ along the $z-$direction \cite{Schall11}. A piezoelectric device was used to move the top boundary in the $ x-$ direction to apply a shear rate of $1.5 \times 10^{-5}$. To prevent boundary-induced crystallization in our samples, the boundaries were coated with a layer of polydisperse particles. Confocal microscopy was used to image the individual particles and to determine their positions in three dimensions with an accuracy of $0.03 \mu m$ in the horizontal and $0.05 \mu m$ in the vertical direction. We tracked the motion of $\sim 2 \times 10^{5}$ particles during a $25$-min time interval by acquiring image stacks every $60~s$. The data was acquired during a small observation window at various strain values $\gamma$.

\subsection{Identification of topological structures using TCC}

The particles are grouped into various topological clusters using the TCC package \cite{Royall13}. The neighbor network of the particles were found using the modified voronoi tesselation. A distance cutoff of $1.3\sigma$ which is the first minima of pair correlation function was used to identify the neighbors. The modified voronoi method uses a four membered ring parameter $f_c$ to remove the long bonds which are not direct neighbors. We used $f_c=0.82$ for our system\cite{Royall13}. The average cluster size and cluster size distribution analysis were done using OVITO\cite{ovito,David15}.

\subsection{Calculation of caging potential }

The caging potential of the particles were found using eq.1. The particle level pair correlation function is mollified using the following expression \cite{Parrinello17_2} 
\begin{equation}\label{eq:3}
g^{i}(r)=\frac{1}{ \rho \mathbf{dr}}\sum_j\frac{1}{\sqrt{2\pi \delta^2}}\text{e}^-\frac{(r-r_{ij})^2}{2\delta^2},
\end{equation}
where $\delta$ is the Gaussian broadening factor that makes local $g^{i}(r)$ continuous. The values of the broadening parameter is $0.02$, and $\mathbf{dr}=4\pi r^2 dr$ respectively. The caging particle is calculated when the particles are in the bulk system and then the caging potential of a particular cluster type is found by only considering the caging potential of those particles.\\

Note that each particle in a cluster experiences a unique local environment, determined by the number and arrangement of its neighboring particles. Since particles may have neighbors from various cluster types, we define central particles within an $n$-particle cluster as those having at least $n-1$ neighbors of the same cluster type. For instance, a particle is classified as part of a $13A$ (regular icosahedral) cluster if it is surrounded by $12$ other particles that also belong to icosahedral clusters. These central particles are expected to exhibit a higher caging potential than the surrounding particles within the same cluster (see Fig.~S2 in the Supplementary Information). The distributions, $P(\beta\phi_i)$, presented in Figs.~1(a) and 2(a) were obtained using the central particle. \\

\subsection{Identifying clusters of particles}
The cluster analysis was performed using an open source application OVITO\cite{ovito}. For the cluster analysis, the first minimum of the pair interaction potential is used as the cutoff distance for identifying connected particles. The mean cluster size is averaged over 20 different snapshots within a strain window of $\delta \gamma=0.0018$, over which no significant structural changes are observed. 

\subsection{Spatial correlation of LFS particles with stable and unstable zones}

The soft particles are defined as those in the bottom $10\%$ of the caging potential distribution, while the hard particles are in the top $10\%$. To examine spatial correlations, we compute the radial pair correlation function between the particles in a given cluster and the particles identified as soft or hard using the expression:

\begin{equation}
g(r) = \frac{V}{4\pi r^2 \Delta r N_\alpha N_\beta} \sum_{i=1}^{N_\alpha} \sum_{j=1}^{N_\beta} \delta(r - \abs{\vec{r}_{ij}}),
\end{equation}

Here, $V$ is the volume of the system. The indices $\alpha$ and $\beta$ refer to LFS and soft/hard particles, respectively. $N_\alpha$ and $N_\beta$ are the respective counts of these particles, and $\abs{\vec{r}_{ij}}$ is the distance between particles $i$ and $j$.

\subsection{Calculation of non-affine displacement of particles}

The non-affine displacement of a tagged particle $i$ over a strain increment of $\Delta\gamma$ is defined as \cite{Langer98, Schall11}
\begin{equation}
D^2_{min,i}(\gamma,\Delta\gamma)=  \frac{1}{N_i}\sum_{j=1}^{N_i} \left[\mathbf{r}^{ij}(\gamma+\Delta \gamma)- \mathbf{\Gamma}_j(\gamma) \mathbf{r}^{ij}(\gamma)\right]^2,
\end{equation}
where $\mathbf{r}^{ij}$ is the displacement vector between particle $i$ and its nearest neighbors $j$, $N_i$ is the number of first nearest neighbors of particle $i$ based on the first minima of $g(r)$, and $\mathbf{\Gamma}_j$ is the best-fit affine deformation tensor that minimizes $D^2_{min,i}$. To get the non-affine displacement of the cluster particles based on their index in the trajectory. 

\section*{Acknowledgements}
R.S. acknowledges Alex Malins, Patrick Royall, Abhinav and Mohit Sharma for helpful discussions. V.C. acknowledges funding from Indian Institute of Science Education and Research Pune as a start-up grant and Department of Science and Technology - Science and Engineering Research Board India under Grant Nos. SRG/2019/001922 and CRG/2021/007824. R.S was supported by a PhD fellowship from CSIR India.

\section*{References}
%

\end{document}